\documentclass[manuscript,twocolumn]{aastex701}

\usepackage{natbib}
\usepackage{hyperref}

\hypersetup{linkcolor=red,citecolor=blue}

\shortauthors{Huang et al.}

\usepackage{graphicx}
\graphicspath{{./figpack/}}
\usepackage{epstopdf}
\usepackage{amsmath}
\usepackage{amssymb}
\usepackage{verbatim}
\usepackage{enumerate}
\usepackage{enumitem}
\setlist[enumerate]{itemsep=0pt, parsep=0pt, topsep=0pt,wide=0pt, leftmargin=0pt} 

\begin{document}

\title{Dual-Perspective Microwave and Hard X-ray Constraints of Asymmetric Nonthermal Loops in an X-class Flare}

\author[orcid=0009-0004-1653-1735]{Ruifei Huang}
\affiliation{Center for Integrated Research on Space Science, Astronomy, and Physics, Institute of Frontier and Interdisciplinary Science, Shandong University, Qingdao 266237, People's Republic of China}
\email{huangrf@mail.sdu.edu.cn}

\author[orcid=0000-0001-6449-8838]{Yao Chen}
\affiliation{Center for Integrated Research on Space Science, Astronomy, and Physics, Institute of Frontier and Interdisciplinary Science, Shandong University, Qingdao 266237, People's Republic of China}
\affiliation{Shandong Key Laboratory of Space Environment and Exploration Technology, Institute of Space Sciences, Shandong University, Weihai 264200, People's Republic of China}
\email[show]{yaochen@sdu.edu.cn}

\author[orcid=0000-0002-7357-211X]{Victor Melnikov}
\affiliation{Central Astronomical Observatory at Pulkovo, Russian Academy of Sciences, Saint Petersburg 196140, Russia}
\affiliation{Center for Integrated Research on Space Science, Astronomy, and Physics, Institute of Frontier and Interdisciplinary Science, Shandong University, Qingdao 266237, People's Republic of China}
\email[show]{vic-meln@mail.ru}

\author[orcid=0000-0001-8644-8372]{Alexey Kuznetsov}
\affiliation{Institute of Solar-Terrestrial Physics, Irkutsk 664033, Russia}
\email{a_kuzn@iszf.irk.ru}

\author[orcid=0000-0002-6985-9863]{Zhao Wu}
\affiliation{Shandong Key Laboratory of Space Environment and Exploration Technology, Institute of Space Sciences, Shandong University, Weihai 264200, People's Republic of China}
\email{wuzhao@sdu.edu.cn}

\author{Dmitriy Smirnov}
\affiliation{N.I. Lobachevsky State University of Nizhny Novgorod, Nizhniy Novgorod 603105, Russia}
\affiliation{Central Astronomical Observatory at Pulkovo, Russian Academy of Sciences, Saint Petersburg 196140, Russia}
\email{dmitriy.smirnov@unn.ru}

\author[orcid=0000-0002-1107-7420]{Sergey Anfinogentov}
\affiliation{Institute of Solar-Terrestrial Physics, Irkutsk 664033, Russia}
\email{anfinogentov@iszf.irk.ru}

\author[orcid=0000-0002-1576-4033]{Feiyu Yu}
\affiliation{Shandong Key Laboratory of Space Environment and Exploration Technology, Institute of Space Sciences, Shandong University, Weihai 264200, People's Republic of China}
\email{feiyuyu@mail.sdu.edu.cn}

\author[orcid=0000-0003-1034-5857]{Xiangliang Kong}
\affiliation{Shandong Key Laboratory of Space Environment and Exploration Technology, Institute of Space Sciences, Shandong University, Weihai 264200, People's Republic of China}
\email{kongx@sdu.edu.cn}

\author[orcid=0000-0001-5483-6047]{Ze Zhong}
\affiliation{School of Astronomy and Space Science and Key Laboratory of Modern Astronomy and Astrophysics, Nanjing University, Nanjing 210023, People's Republic of China}
\email{zezhong@sdu.edu.cn}

\author[orcid=0000-0003-4956-6040]{Mingzhe Guo}
\affiliation{Center for Integrated Research on Space Science, Astronomy, and Physics, Institute of Frontier and Interdisciplinary Science, Shandong University, Qingdao 266237, People's Republic of China}
\email{m.guo@email.sdu.edu.cn}

\author[orcid=0000-0001-8132-5357]{Hao Ning}
\affiliation{Center for Integrated Research on Space Science, Astronomy, and Physics, Institute of Frontier and Interdisciplinary Science, Shandong University, Qingdao 266237, People's Republic of China}
\email{haoning@sdu.edu.cn}

\correspondingauthor{Yao Chen, Victor Melnikov}

\begin{abstract}
We report dual-perspective microwave and HXR observations of an X-class flare on 2024 May 15, taking advantage of the unique geometry of a front-side view from the Earth and a back-side perspective from the Solar Orbiter (SolO). Using spatially resolved imaging spectroscopy from Siberian Radioheliograph (SRH) together with Chashan Broadband Solar millimeter spectrometer (CBSmm) and STIX data, we identify a set of nonthermal flaring loops in an asymmetric magnetic field, with microwave sources located near the loop top and HXR sources associated with the southern footpoint. Compared with HXR, the microwave emission shows an opposite ascending trend, an increasing time lag in time profile, and a distinctive ``SHH'' spectral pattern, which we attribute to energy-dependent trapping and precipitation of energetic electrons in an asymmetric magnetic configuration. Flux pulsations and their spectral and polarization signatures are consistent with intermittent particle acceleration rather than MHD wave modulation. 
Microwave magnetic diagnostics, corroborated by non-linear force free field (NLFFF) extrapolation, provide key constraints on the three-dimensional magnetic configuration. The dual-perspective flux profile comparison and consistent QPP signatures across wavelengths together support a self-consistent picture of energy-dependent electron trapping, precipitation, and transport in these asymmetric loops.

\end{abstract}


\section{Introduction}

Nonthermal microwave and HXR emissions carry crucial information about particle acceleration and transport processes during solar flares \citep{Ramaty1969,Dulk1985,White2011}. Both emissions originate from the same group of nonthermal energetic electrons but differ in radiation mechanisms, energy domains, and atmospheric layers. Microwave emission arises primarily from gyrosynchrotron radiation produced by electrons with energies ranging from hundreds of keV to several MeV gyrating in coronal magnetic fields, while HXR emission results mainly from bremsstrahlung radiation produced when tens to hundreds of keV electrons collide with the dense plasma at footpoints. 

Nonthermal observations from multiple perspectives are crucial for inferring the three-dimensional (3D) magnetic configuration of solar flares. The Advanced Space-based Solar Observatory \citep[ASO-S;][]{Gan2019} and Solar Orbiter \citep[SolO;][]{Muller2020} have both carried HXR detectors. Their coordinated observations enable stereoscopic detection of nonthermal processes in solar flares from two vantage points. Combined with microwave observations from ground-based telescopes represented by the Siberian Radioheliograph \citep[SRH;][]{Altyntsev2020}, Expanded Owens Valley Solar Array \citep[EOVSA;][]{Gary2018}, and Mingantu Spectral Radioheliograph \citep[MUSER;][]{Yan2021}, this multi-perspective, multi-wavelength approach constrains the physical characteristics of flaring structures and processes. 
Studies along this line remain relatively rare. \citet{ryan2024} combined HXR data from ASO-S and SolO with a small separation angle to conduct 3D triangulation of a solar flare and its loop structure. \citet{Collier2024} utilized EOVSA and SolO to identify microwave and HXR synchronous pulsations with shifting periods and spatially separated sources along flare ribbons. \citet{kaltman2026} derived 3D diagnostics of magnetic field strength, Alfvén speed, and plasma beta in a flaring volume using microwave and HXR images taken from two vantage points by EOVSA, SolO and Hinode \citep{Kosugi2007}, providing constraints for modeling the magnetic reconnection and flare dynamics.

Quasi-periodic pulsations (QPPs) in microwave and HXR fluxes are commonly observed during solar flares \citep[e.g.,][]{Melnikov2005,Zimovets2009,Tan2016,Hayes2019,Li2022,Inglis2023}, with characteristic periods ranging from sub-second to minutes \citep[e.g.,][]{Inglis2009,Kupriyanova2010,Li2020}. 
Two major groups of mechanisms have been proposed to explain QPPs. The first group involves direct modulation by magnetohydrodynamic (MHD) waves propagating in coronal loops, making magnetic field strength, viewing angle, and plasma density vary periodically \citep[e.g.,][]{Nakariakov2003,Kim2012,Kolotkov2015,Tian2016}. The second group involves intermittent energy release processes such as magnetic reconnections, according to which the observed periodicity may result from the statistical properties of successive energy release processes \citep{Nakariakov2009,VanDoorsselaere2016,McLaughlin2018,Kupriyanova2020,Zimovets2021}.

In this paper, we analyze joint observations from Earth-based and backside perspectives of an X3.5-class solar flare exhibiting flux pulsations in both microwave and HXR emissions. Since the STIX imaging position is uncertain for this event, our analysis does not rely on direct HXR source localization. Instead, we combine microwave imaging spectroscopy, temporal microwave–HXR correlations, and magnetic-field modeling, to diagnose the overall magnetic structure, infer the asymmetric transport pattern of energetic electrons, and constrain the physical origin of the observed pulsations.
We overview the event in Section~\ref{Sect.2}. Section~\ref{Sect.3} presents detailed analyses of the pulsations, including their temporal characteristics, spectral dynamics, source localization, and magnetic diagnostics derived from our spatially resolved fittings. Section~\ref{Sect.4} discusses the underlying physics and Section~\ref{Sect.5} summarizes our results.

\section{Event Overview from Two Perspectives}\label{Sect.2} 

The X3.5-class solar flare occurs on 2024 May 15 in active region (AR) 13664 at (S18, W89) viewed from the Earth. GOES soft X-ray (SXR) data show that the flare commences at 08:18 UT and peaks at 08:37 UT. A halo coronal mass ejection (CME) accompanies this flare, with a linear speed of 1648 km/s according to SOHO/LASCO C2 \citep{Brueckner1995}.

The flare is observed by a comprehensive suite of space-based instruments, including the Atmospheric Imaging Assembly \citep[AIA;][]{Lemen2012}, Helioseismic and Magnetic Imager \citep[HMI;][]{Scherrer2012} onboard the Solar Dynamics Observatory \citep[SDO;][]{Pesnell2012}, and the Extreme Ultraviolet Imager \citep[EUI;][]{Rochus2020}, Polarimetric and Helioseismic Imager \citep[PHI;][]{Solanki2020}, Spectrometer Telescope for Imaging X-rays \citep[STIX;][]{Krucker2020} onboard SolO. It is also recorded by ground-based facilities, including SRH which comprises three separate arrays covering 3 to 24 GHz through 48 frequency channels, with a temporal cadence of 3.5 s, and the Chashan Broadband Solar millimeter spectrometer (CBSmm; \citealt{Shang2022,Shang2023,Yan2023}) at 35--40 GHz with a high temporal resolution of $\sim$ 0.5 s. These instruments provide broadband multiwavelength coverage, from HXR, to extreme ultraviolet (EUV), and microwaves.

\begin{figure}[htbp]
	\centering
	\includegraphics[width=0.95\textwidth]{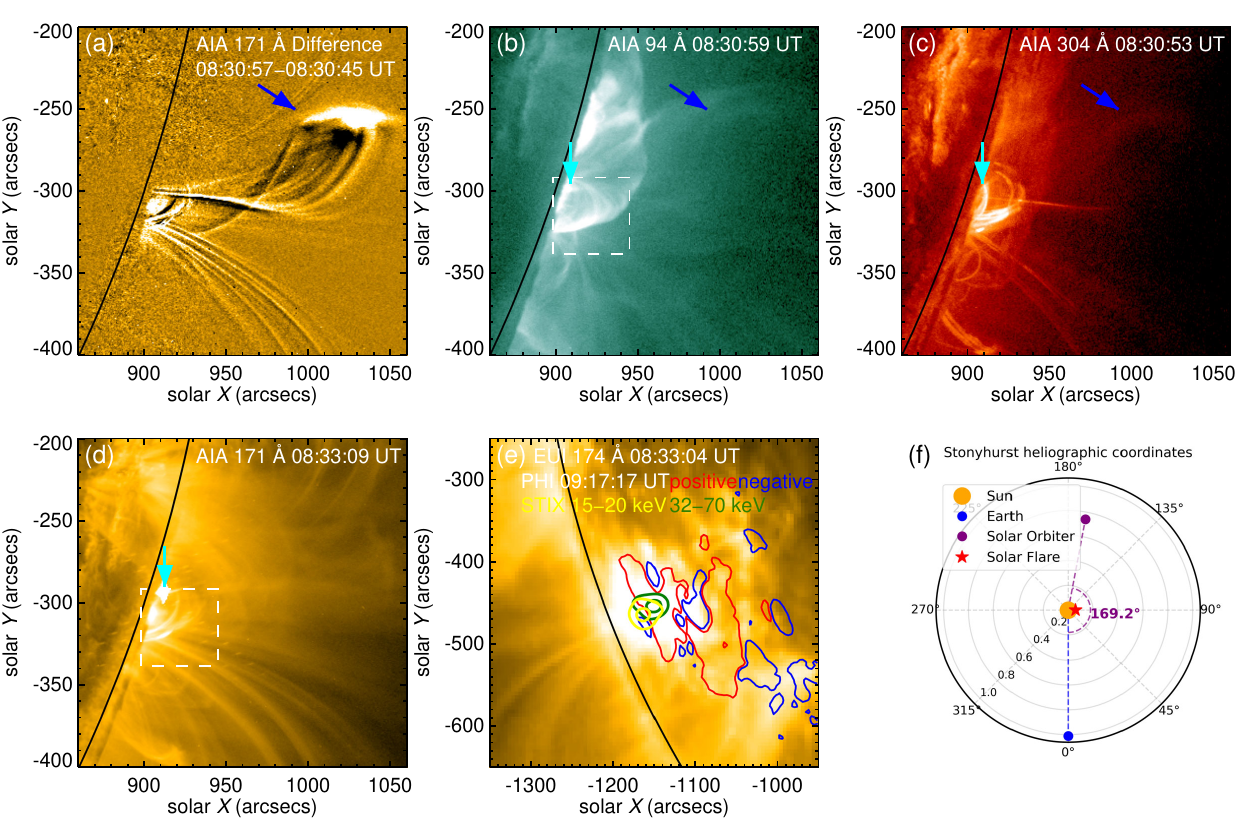}
	\caption{Overview of the X3.5-class flare on 2024 May 15. (a) SDO/AIA 171~\AA\ running-difference image at 08:30:57 UT. (b)--(d) AIA 131, 304, 171~\AA\ images at 08:30:59, 08:30:53 and 08:33:09 UT. Blue arrows in (a)--(c) denote the twisted ejecta. White boxes in (b) and (d) outline the field of view in \ref{figure7}(a). (e) SolO/EUI 174~\AA\ image of the same active region at 08:33:04 UT, overlaid with red and blue contours at $\pm 120$ G of PHI line-of-sight magnetogram at 09:17:17 UT. Yellow and dark green contours given by 30\%, 75\% of the maximum intensity represent the STIX HXR sources without manual shift at 15--20, 32--70 keV. Black lines in (a)--(d) depict the solar limb. (f) Relative positions of the Sun, Earth and SolO. Red star shows the position of this flare. All times are presented in UTC at the Earth. An animation of this figure showing the evolution of SDO/AIA 171~\AA\ running-difference and 131, 171, 304~\AA\ images from 08:29 UT to 08:38 UT is available.}
	\label{figure1}
\end{figure}

Figure~\ref{figure1} displays the event overview. SolO observes this event from the far side of the Sun at a distance of $\sim 0.74\,$AU and an angle of $\sim 169.2^\circ$ relative to the Earth-Sun line (Figure~\ref{figure1}(f)). Thus, SDO and SolO witness the flare from their western and eastern solar limbs, respectively. SolO/EUI is operated at a cadence of 10 minutes and the SolO/PHI data closest in time is $\sim 45$ minutes later. A light travel time lag of $\sim 134.2\,$s exists between the two vantage points. We applied temporal correction to all SolO data throughout our analysis and presented all times in UTC at the Earth.

As shown in Figure~\ref{figure1}(e), the AR is in a complex multipolar magnetic configuration and completely rotates to the back side according to the line-of-sight magnetogram from SolO/PHI. From the SDO (or the Earth) perspective, the flare is partially occulted. The accompanying animation showing the EUV evolution suggests that ejecta observed in high and low temperature bands (indicated by blue arrows in Figure~\ref{figure1}(a)--(c)) rises and marks the onset around 08:24 UT. Later its legs evolve into an X-shaped structure, beneath which are bright loops moving sunward. After 08:30 UT, a collimated EUV ray-like structure appears to the north of the loops and moves northward at a speed of $\sim$ 19 km/s (see the cyan arrows in Figure~\ref{figure1}(b)--(d)). From the SolO perspective, STIX observes single HXR sources at 15--20 and 32--70 keV and are cospatial with these bright flaring loops (shown in Figure~\ref{figure1}(e) without manual shift). 
\section{Analysis of Multiwavelength Data}\label{Sect.3} 

In this section, we analyzed microwave and HXR emissions from the two perspectives (the Earth and SolO/STIX), presenting their spatiotemporal and spectral characteristics, then conducted magnetic diagnostics.

\subsection{Temporal and Spectral Characteristics}\label{Sect.3.1} 

\subsubsection{Pulsations of Microwave and HXR Flux Curves}\label{Sect.3.1.1} 

\begin{figure}[htbp]
	\centering
	\includegraphics[width=0.7\textwidth]{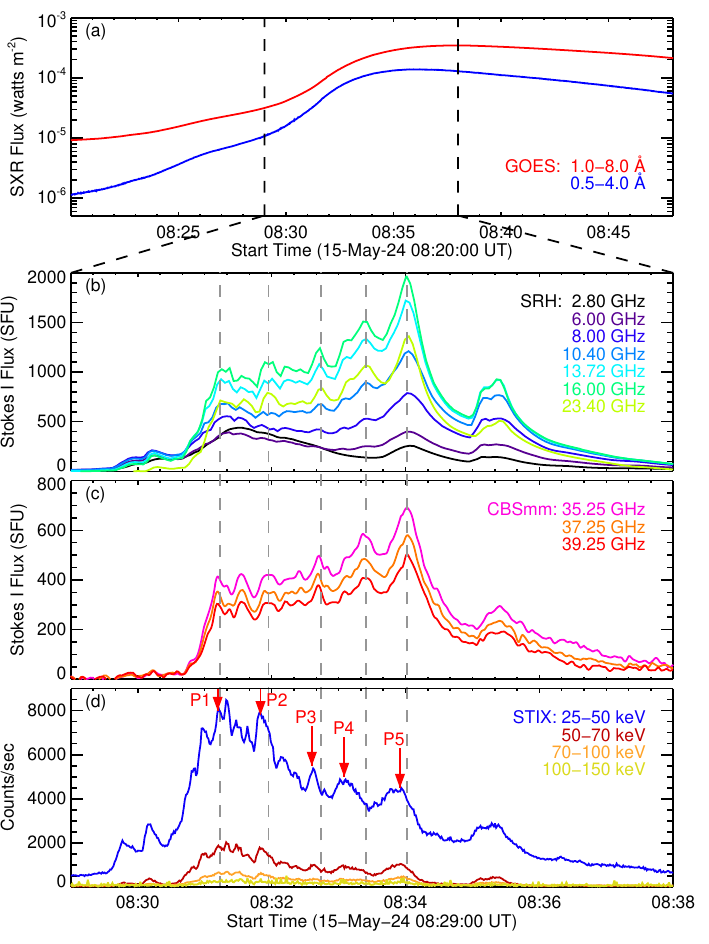}
	\caption{(a)--(d) Flux curves in SXR, 3--24 GHz, 35--40 GHz and 25--150 keV from GOES, SRH, CBSmm and SolO/STIX, respectively. All times are presented in UTC at the Earth. Gray vertical dashed lines in (b)--(d) indicate the five major microwave peaks at 08:31:13, 08:31:58, 08:32:45, 08:33:24 and 08:34:00 UT. P1--P5 in (d) mark the five major HXR peaks, which slightly precede the former.} 
	\label{figure2}
\end{figure}

\begin{figure}[htbp]
	\centering
	\includegraphics[width=0.8\textwidth]{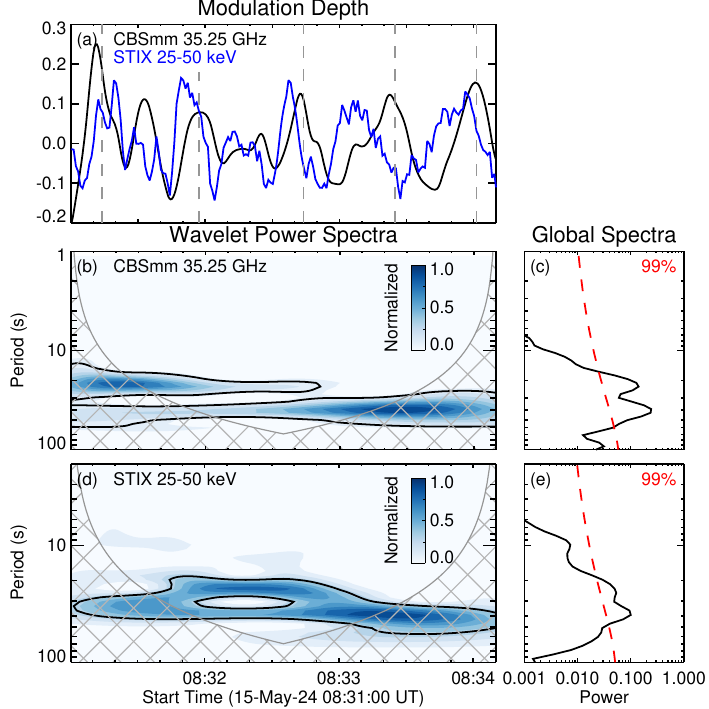}
	\caption{(a) Modulation depths of flux density at 35.25 GHz (black) from CBSmm and 25--50 keV (blue) from STIX. Gray vertical dashed lines indicate the five major microwave peaks shown in Figure~\ref{figure2}. (b)--(e) Normalized wavelet power spectra and their global spectra of the modulation depths. Red dashed lines in (c) and (e) indicate the 99\% significance level.}
	\label{figure3}
\end{figure}

Figure~\ref{figure2} shows the total (i.e., spatially unresolved) microwave and HXR flux curves. In the impulsive phase, microwave emissions at $\sim$ 10--24 GHz and 35--40 GHz exhibit in-phase oscillations persisting for $\sim$ 3 minutes (08:31--08:34 UT), with five major peaks (denoted by gray dashed lines in Figure~\ref{figure2}(b)--(d)) that become more pronounced as frequency increases. These high-frequency curves show an overall ascending trend, while curves at 2.8--7.6 GHz remain smooth with only two major peaks. HXR emissions at 25--50 keV and 50--70 keV display overall decreasing profiles, yet peaks ahead of the microwaves with increasing separation up to $\sim$ 20 s (see the comparison of P1--P5 and gray dashed lines in Figure~\ref{figure2}(d), and curves in Figure~\ref{figure3}(a)).

Following the wavelet method \footnote[1]{\url{http://atoc.colorado.edu/research/wavelets/}} employed by \citet{Torrence1998}, we first smoothed CBSmm flux curve to the SRH resolution ($\sim$3.5 s), then deduced the modulation depth of detrended microwave and HXR profiles by applying smoothing windows varying from 10 to 60 s. The wavelet analysis presented in Figure~\ref{figure3} shows two periods of $\sim$ 39 s and $\sim$ 25 s in curves at $>$ 10 GHz and 25--50 keV. The fact that the same oscillating periods appear in both microwave and HXR emissions suggests that they may have a common origin.

\subsubsection{Spectral Fitting of Total Fluxes}\label{Sect.3.1.2} 

\begin{figure}[htbp]
	\centering
	\includegraphics[width=0.95\textwidth]{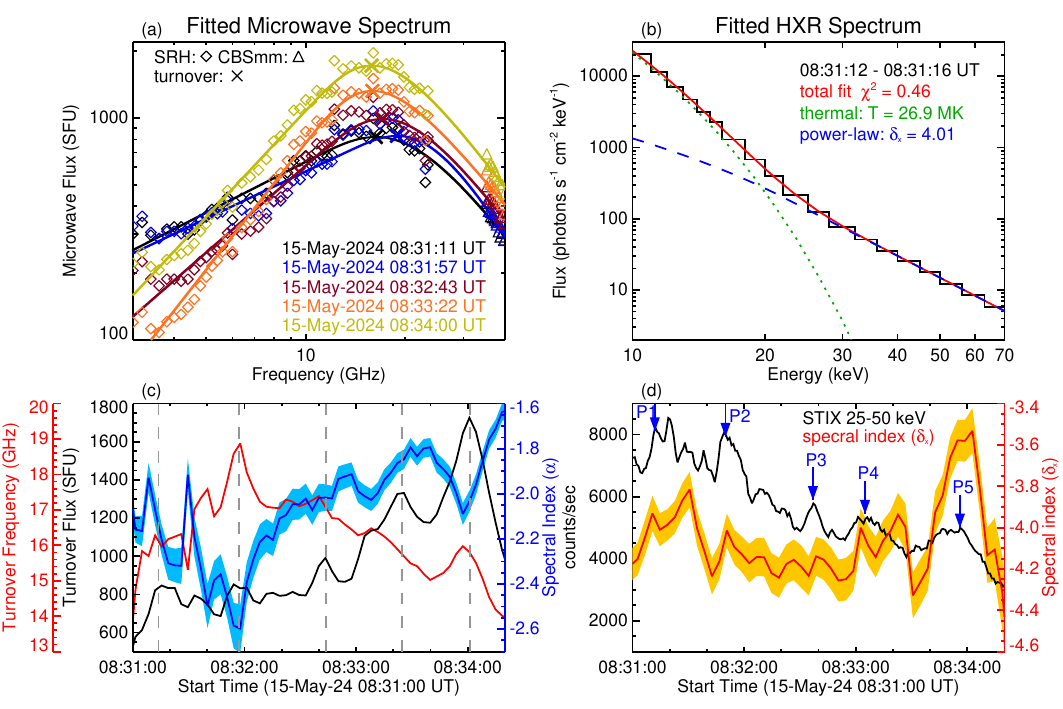}
	\caption{(a) Fitted total gyrosynchrotron spectra of SRH (diamonds) and CBSmm (triangles) at the five major peaks in Figure~\ref{figure2}(b) and (c). Crosses mark turnover frequencies. (b) STIX HXR spectrum fitted with thermal (green dotted) and thick-target bremsstrahlung (blue dashed) models for 08:31:12--08:31:16 UT. Red and black lines show the total fitting and STIX data with background subtracted. The fitting chi-square value is $\chi^2=0.46$. (c) Temporal evolution of fitted microwave spectral index $\alpha$ in optically-thin regime (blue), turnover frequency (red) and turnover flux (black). Gray vertical dashed lines indicate the five major microwave peaks shown in Figure~\ref{figure2}. (d) Evolutions of STIX 25--50 keV flux (black) and fitted electron spectral index $\delta_x$ (red). Light blue and orange bands in (c) and (d) represent $1\sigma$ errors of $\alpha$ and $\delta_x$, respectively.}
	\label{figure4}
\end{figure}

To study the evolution of spectral characteristics, we used SRH and CBSmm data to fit the total microwave spectra during the pulsations (see Figure~\ref{figure4}(a) and (c)) and obtained typical gyrosynchrotron features. 
From 08:31:00 to 08:36:40 UT, the fitted spectral index ($\alpha$) in the optically-thin regime varies from $\sim$ -2.6 to -1.6, manifesting overall hardening with multiple peaks. The $\alpha$ peak occurs progressively later than the corresponding flux peak, forming a distinctive pattern of ``soft---hard---harder (SHH)''.
The turnover frequency rises rapidly at $\sim$ 08:31:00 UT and reaches a maximum of 18.9 GHz one minute later. It maintains high values $>$ 14 GHz throughout the pulsations and peaks with the flux. The profile of flux at turnover frequency resembles that at $>$ 10 GHz shown in Figure~\ref{figure2}(b) and (c). These features give a positive correlation among turnover frequency, turnover flux, and spectral index, providing evidence for the origin of high turnover frequencies.

We further used the thermal (\texttt{f\_vth.pro}) and thick-target bremsstrahlung (\texttt{f\_thick2.pro}) models in the standard OSPEX software package \citep{Tolbert2020} to fit the HXR spectra, over an energy range of 10--70 keV and an integral time of 4 s. Figure~\ref{figure4}(b) shows the fitting at P1 as an example. Figure~\ref{figure4}(d) displays the evolution of fitted spectral index of energetic electrons ($\delta_x$) and 25--50 keV flux curve. The index $\delta_x$ varies from $-4.2$ to $-3.5$, synchronously with the flux curve. It reaches local maxima at times close to the flux peaks, displaying the usual ``soft---hard---soft (SHS)'' feature. We highlight the difference between the ``SHH'' pattern in microwave emission and the ``SHS'' pattern in HXR emission, which indicates the electron transport processes in flaring loops, which will be discussed in Section~\ref{Sect.4}.

\subsection{Microwave Sources}\label{Sect.3.2} 

\begin{figure}[htbp]
	\centering
	\includegraphics[width=0.95\textwidth]{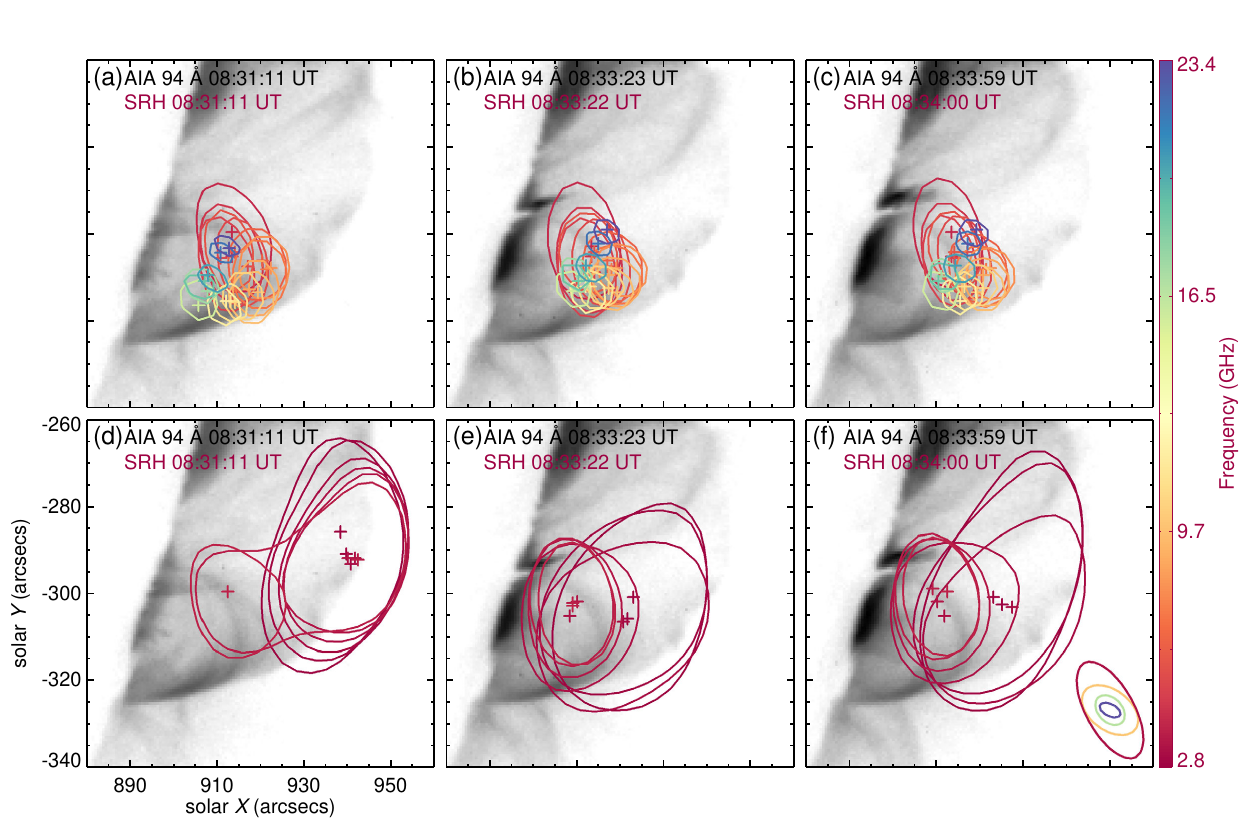}
	\caption{(a)--(c) Microwave sources at 4.2--23.4 GHz superimposed onto AIA 94~\AA\ images (shown in reverse grayscale) at 08:31:11, 08:33:22, 08:34:00 UT, i.e., the first, fourth and fifth microwave peaks. Contours are 90\% of maximum brightness temperature, whose position is marked by crosses at each frequency. (d)--(f) Same as (a)--(c) but for 2.8--4.0 GHz. SRH beam sizes at 3.4, 6.0, 16.76, 23.40 GHz are indicated in the bottom right of (f). An animation of this figure showing the evolution of SRH sources overlaid on AIA 94~\AA\ images from 08:29 UT to 08:34 UT is available.}
	\label{figure5}
\end{figure}

To localize the microwave pulsations and investigate their spatial association with EUV structures, we examined SRH imaging data from 2.8 to 23.4 GHz. According to Figure~\ref{figure5} and the accompanying animation showing its temporal evolution, microwave sources at $\nu>4$ GHz lie around the loop top throughout the pulsations, while sources at $\nu \le4$ GHz first move anti-sunward with increasing flux, then sunward with declining flux. No strong microwave sources are detected in lower atmosphere, suggesting that the nonthermal electrons inducing microwave emission are primarily confined to the upper part of the loops.

\begin{figure}[htbp]
	\centering
	\includegraphics[width=0.95\textwidth]{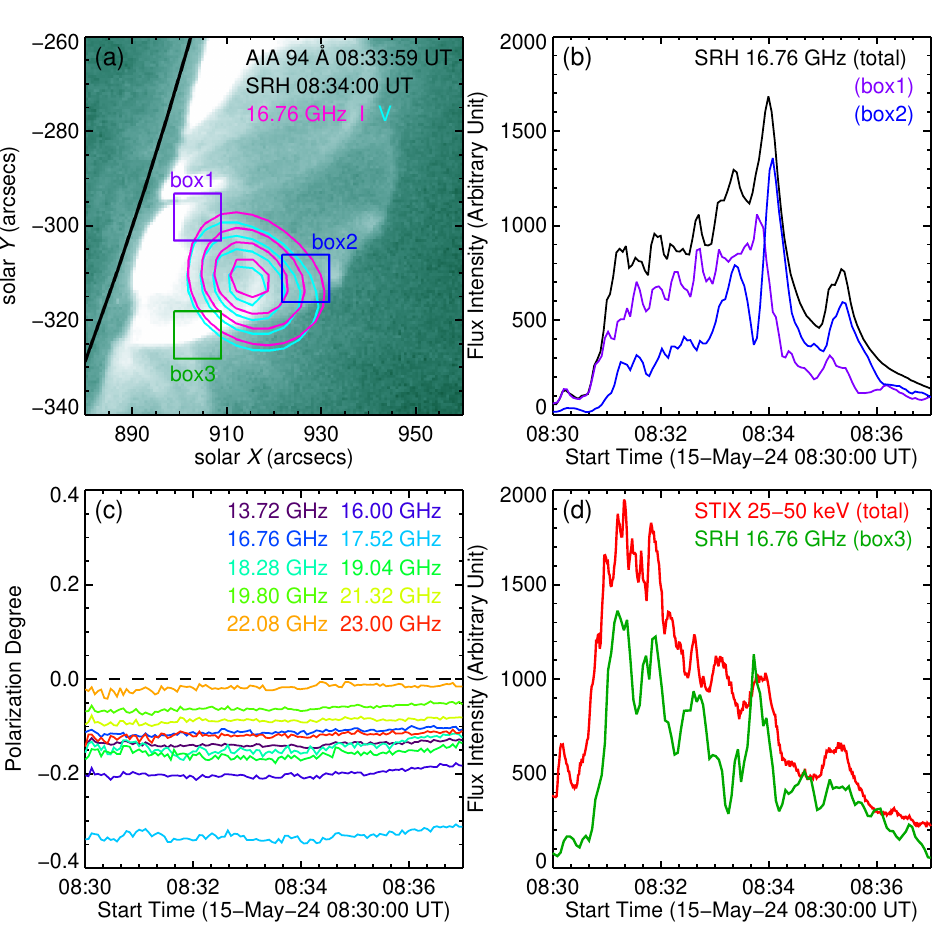}
	\caption{(a) Stokes $I$ (magenta) and $V$ (cyan) sources at 16.76 GHz superimposed onto AIA 94~\AA\ image at 08:34:00 UT. Contours are 30\%, 50\%, 70\%, 90\% of the maximum brightness temperature. Boxes 1--3 (3$\times$3 pixels) represent the northern loop leg, loop top and southern loop leg. (b) Total flux profiles at SRH 16.76 GHz (black) and its integrations in boxes 1 and 2 (purple, blue). (c) Temporal evolution of microwave polarization degrees at $\sim$ 14--23 GHz. (d) Same as (b) but for total flux profiles at STIX 25--50 keV (red) and integration at SRH 16.76 GHz in box 3 (green).}
	\label{figure6}
\end{figure}

We integrated microwave flux over the northern, top and southern parts of the loop set (denoted by boxes 1--3 in Figure~\ref{figure6}(a), respectively). As shown in Figure~\ref{figure6}(b) and (d), curves from the top and northern leg resemble the total microwave flux, with similar oscillation pattern and overall increasing trend, whereas the microwave flux in the southern leg displays an overall declining trend, with peaks preceding those of the total microwave flux. This is a key result. Note that the similarity is based on flux profiles rather than imaging positions, and establishes a physical connection between the HXR-emitting electrons and the southern leg of the loop system.

Figure~\ref{figure6}(a) also displays the Stokes $V$ image which is cospatial with Stokes $I$ map. We derive the background-subtracted circular polarization degree in optically-thin regime, which reaches $\sim-40\%$ at 17.52 GHz and remains stable throughout the QPPs (see Figure~\ref{figure6}(c)). Since polarization is highly related to the variation of magnetic field, it provides important constraints on the nature of the QPP driver.

\subsection{Magnetic Diagnostics}\label{Sect.3.3} 

To further determine the magnetic configuration of this flare, we applied the homogeneous gyrosynchrotron model in GSFIT package \citep[][]{Fleishman2020,Fleishman2022} to fit observed spectrum at each pixel.
We also employed its method to reduce the impact of beam size variation \citep[c.f.,][]{gary2013}. We assumed a single isotropic power-law spectrum for energetic electrons of 10 keV to 10 MeV, fixed source size (area $3.3^{\prime\prime}\times 3.3^{\prime\prime}$, depth $8^{\prime\prime}$) for each pixel, and plasma temperature $T =10^7$~K. 
Other free parameters include the magnetic field strength ($B$), viewing angle ($\theta$, i.e., the angle between magnetic field and line of sight), number density of nonthermal electrons ($n_t$) with energies $E>10$ keV, energy spectral index ($\delta$), and ambient plasma density ($n_0$).

\begin{figure}[htbp]
	\centering
	\includegraphics[width=0.95\textwidth]{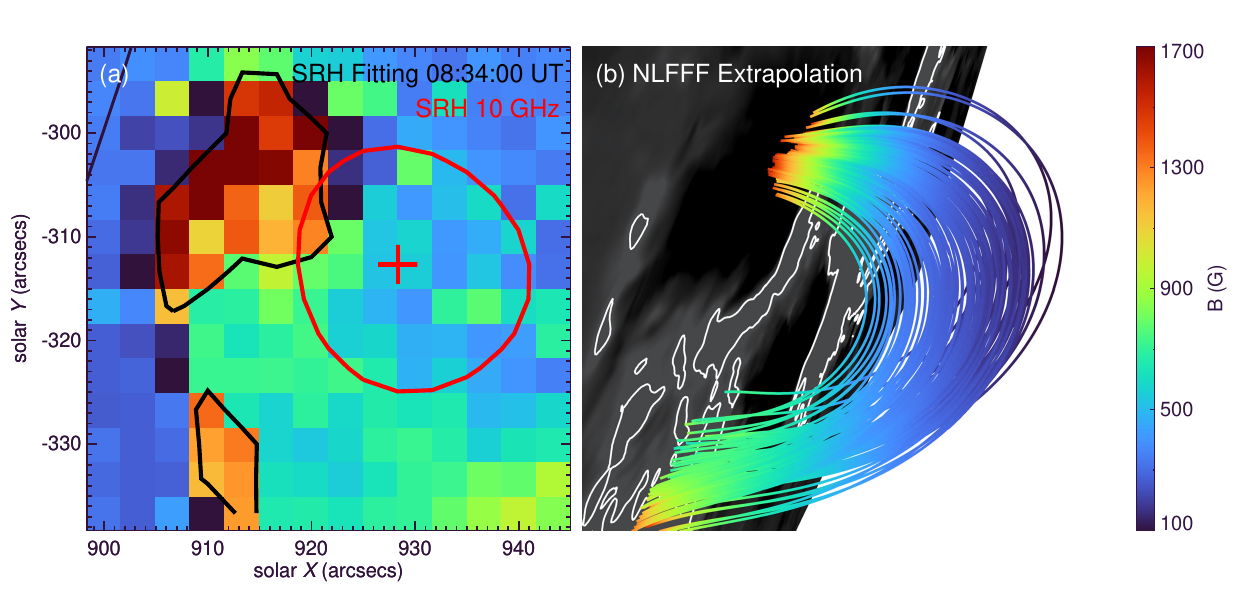}
	\caption{(a) Magnetic field distribution from microwave spectral fitting at 08:34:00 UT, within the white box shown in Figure~\ref{figure1}(b) and (d). Black and red contours outline the 56\% of the maximum field value and maximum brightness temperature at 10 GHz, respectively. Red cross indicates the position of the brightness maximum of the source at 10 GHz. (b) Selected field lines colored by field strength from the NLFFF extrapolation. Background is SDO/HMI magnetogram at 02:00 UT on 2024 May 12 with white contours showing $700$ G polarities.}
	\label{figure7}
\end{figure}

Figure~\ref{figure7}(a) presents the obtained $B$ map and its correlation with microwave source at 10 GHz that demonstrates the position of loop top. Particularly, the $B$ map exhibits an asymmetric loop structure: $B$ reaching $\sim1600\pm100$ G at the northern part of the loops, $\sim1300\pm100$ G at the southern part, and $\sim700\pm100$ G around the loop top. This implies a unique magnetic geometry, which affects the transport of energetic electrons along the flaring loops.  

Both SDO/HMI and SolO/PHI indicate that the overall large-scale polarity distribution persists for days. We therefore used the SDO/HMI magnetogram at 02:00 UT on 2024 May 12 that is three days earlier, to extrapolate the magnetic field morphology via the non-linear force free field (NLFFF) technique \citep{Guo2016} (shown in Figure~\ref{figure7}(b)). From the extrapolation, we identified similar asymmetric loops connecting the stronger northern polarity and the weaker southern counterpart. This independently supports the asymmetric magnetic picture and demonstrates the diagnostic power of microwave spectral imaging for constraining coronal magnetic fields.

\section{Discussion on the Physical Scenario}\label{Sect.4} 

\subsection{Localization of HXR Sources and Inference from Flux Profile Similarity}  

Direct imaging localization of the nonthermal HXR sources from STIX is not feasible for this event. The source positioning uncertainty can reach to tens of arcsecs due to limited accuracy of STIX aspect system \citep[e.g.,][]{Massa2022,Collier2024,Bajnokova2024}. The low cadence (10 minutes) and severe saturation of EUI images preclude a reliable alignment. Data from ASO-S/HXI are also unusable owing to intense particle background. 

However, the similarity between the microwave flux profile from the southern leg and HXR 25--50 keV total flux from STIX (shown in Figure~\ref{figure6}(d)) provides an independent diagnostic and suggests a physical connection between the HXR-emitting electrons and the southern leg of the loop system. We therefore infer that the nonthermal HXR sources are likely located at the southern footpoint.

\subsection{Physical Self-Consistency of the Asymmetric Loop Scenario}

This southern-footpoint inference of nonthermal HXR sources is physically consistent with our spatially-resolved spectral fittings, which reveal an asymmetric magnetic configuration in the flaring loops. This results in preferential precipitation of energetic electrons in the southern part where $B$ is weaker \citep[e.g.,][]{Yu2026}.

This scenario provides a unified interpretation of the main observational features. The flux profiles in microwave at $\nu>10$ GHz and HXR 25--50 keV exhibit contrary overall trends with an increasing lag at flux peaks with time. Microwave spectra show a ``SHH'' pattern, rather than the typical ``SHS'' pattern in HXR emission.
These can be interpreted in terms of energy-dependent transport effects in the asymmetric loops, and different sensitivities of microwave and HXR emissions to the energy bands of radiating electrons \citep[e.g.,][]{Melnikov1994,Melnikov1998b}.
The lifetime of electrons depends on energy and is defined by Coulomb collision rate: $\tau(E) = A E^{3/2} / n_e$, so electrons with higher energy live longer. In addition, higher-energy electrons are preferentially retained in the loop-top trap, while lower-energy electrons dissipate more rapidly. The former accounts for the loop top microwave source and the latter for the HXR emission. 

The HXR time profile reflects the overall injection rate, while the microwave profile reflects the instantaneous high-energy electron population within the source. The overall declining HXR flux may result from less injection for each pulse, while each new injection adds more high-energy electrons leading to more high-energy electrons in the loop top and rising microwave light curve there. This process hardens the microwave spectrum with time, explains its ``SHH'' pattern, and leads to the increasing microwave–HXR time lag.
Eventually, the magnetic asymmetry, the inferred HXR location, and the contrasting microwave–HXR temporal and spectral behaviors form a self-consistent physical picture.

\subsection{The Origin of QPPs and Energy Release Process}
Several observational features argue against MHD wave modulation as the origin of the observed QPPs and instead favor intermittent particle acceleration/injection. 
First, the microwave and HXR emissions share similar periods, indicating that both emissions are produced by the same population of energetic electrons. 
Second, both emissions exhibit flux pulsations superposed with minor peaks and repeated spectral hardening, which are characteristic of multiple acceleration-injection episodes that directly modulate the radiating electron population. 
Third, the oscillations in the microwave optically-thin and optically-thick regimes are synchronous without a phase difference, which is not expected in simple MHD models for isotropic sources \citep{Mossessian2012,Kuznetsov2015}. 
Fourth, the circular polarization degree in the microwave optically-thin regime remains stable throughout the pulsations, making viewing-angle modulation by MHD waves less favorable \citep{Dulk1985}. 
Besides, the quasi-periodic features do not appear in flux at 2.8--6 GHz. It reflects the relative insensitivity of optically-thick emission to changes in the energetic electron population.

We also observed high spectral turnover frequencies ($>$14 GHz) throughout the pulsations, together with a positive correlation among the turnover frequency, turnover flux, and spectral index ($\alpha$) in the optically-thin regime (shown in Figure~\ref{figure4}(b)). This points to self-absorption, rather than Razin suppression, as the origin of the high turnover frequencies \citep[][see also \citealp{Yan2023,Wu2024} for latest studies]{Twiss1954,Razin1960,Melnikov2008}.

The collimated EUV ray-like structure observed north of the loops may represent a current sheet as the energy release site and link the erupting ejecta and the main flaring region. Although the intrinsic 3D structure and projection effects of this limb event prevent a detailed identification, we suggest that energetic electrons, after intermittent acceleration near this current sheet, are injected or transported into the asymmetric loop system and subsequently trapped there. Rapid shrinkage of reconnected field lines, direct injection, and transverse diffusion may all contribute to this transport process.

\section{Summary}\label{Sect.5} 

We have presented dual-perspective observations of asymmetric nonthermal flaring loops in an X3.5 flare on 2024 May 15, using the unique geometry of a front-side view from Earth and a back-side perspective from Solar Orbiter. Microwave imaging from SRH reveals loop-top sources, while SolO/STIX provides back-side HXR coverage. Microwave spectral fitting indicates the magnetic asymmetry, with $\sim1600\pm100$ G at the northern footpoint, $\sim1300\pm100$ G at the southern footpoint, and $\sim700\pm100$ G at the loop top, supported by NLFFF extrapolation. We inferred that the HXR emission is associated with the weaker-field southern footpoint. The microwave and HXR pulsations share periods of $\sim39$ s and $\sim25$ s, yet display an increasing mutual time lag reaching $\sim20$ s and opposite intensity trends—consistent with energy-dependent electron trapping and precipitation in the asymmetric loops. 

Regarding the origin of these pulsations, several features argue against MHD wave modulation: the in-phase oscillations between microwave and HXR emission intensities, and between optically-thin and optically-thick regimes, the repeating hardening microwave spectra, and the stable circular polarization degree throughout the impulsive phase. All these features are consistent with intermittent particle acceleration, plausibly near a current sheet connecting the erupting ejecta and the flaring region, as the more likely driver. The high turnover frequencies and their correlation with the evolution of spectral index further support self-absorption as its dominant mechanism.

Our microwave magnetic diagnostics, supported by NLFFF extrapolation, provide quantitative constraints on the asymmetric three-dimensional magnetic configuration. These results demonstrate how dual-perspective microwave–HXR constraints can independently establish the magnetic topology and energy-dependent precipitation patterns that govern flare electron dynamics.

\begin{acknowledgments}
This work is supported by grants of National Natural Science Foundation of China 42127804, 42574218, and 42561160095, the National Key R\&D Program of China 2022YFF0503002, and Shandong Provincial Natural Science Foundation ZR2025MS88. A.K. was supported by the Ministry of Science and Higher Education of the Russian Federation, and Chinese Academy of Sciences President's International Fellowship Initiative, PIFI Group grant No. 2025PG0008. D.S. was supported by the Russian Science Foundation grant no. 22-12-00308-P. The authors are grateful to the GOES, SDO, SRH, CBSmm and SolO consortia for providing the data and Muriel Stiefel from ETH Zurich and FHNW for processing STIX data. 
\end{acknowledgments}


\bibliographystyle{aasjournal}
\bibliography{references}


\newpage

\end{document}